\documentclass[onecolumn,a4]{aa}
\usepackage{natbib}
\bibpunct{(}{)}{;}{a}{}{,}
\usepackage{graphicx}
\usepackage{txfonts}
\usepackage[hang,bf,nooneline]{subfigure}
\usepackage[hang,bf,nooneline]{caption}
\usepackage{rotating}

\begin{document}

\title{A new Determination of the Extragalactic Background of Diffuse Gamma Rays
taking into account Dark Matter Annihilation}

\author{W. de Boer, A. Nordt, C. Sander, V. Zhukov}

\institute{Institut f\"ur Experimentelle Kernphysik, Universit\"at Karlsruhe (TH), P.O. Box 6980, 76128 Karlsruhe, Germany\\
\email{Wim.de.Boer@cern.ch, sander@ekp.uni-karlsruhe.de, zhukov@ekp.uni-karlsruhe.de}}

\date{Received November 30, 2006; accepted May 2, 2007}

\authorrunning{de Boer \textit{et al.}}
\titlerunning{Extragalactic Background of Diffuse Gamma Rays}

\abstract{The extragalactic background (EGB) of diffuse gamma rays can be
determined by subtracting the Galactic contribution from the data. This requires a
Galactic model (GM) and we include for the first time the contribution of dark
matter annihilation (DMA), which was previously proposed as an explanation for the EGRET excess
of diffuse Galactic gamma rays above 1 GeV.

In this paper it is
shown that the newly determined EGB  shows a characteristic high energy bump
on top of a steeply falling soft contribution. The bump is shown to be compatible with a
contribution from an extragalactic DMA signal  from weakly interacting massive
particles (WIMPs) with a mass between 50 and 100 GeV in agreement with the
EGRET excess of the Galactic diffuse gamma rays and in disagreement with earlier
analysis.
The remaining soft contribution  of the EGB is
shown to resemble the spectra of the observed point sources in our Galaxy.

  }

\maketitle

\section{Introduction}

In a previous paper (\citet{deboer}, in the following called Paper I, we discussed
the possibility to explain the observed excess in the EGRET data \citep{hunter} in
the diffuse Galactic gamma rays above 1 GeV as a dark matter annihilation (DMA)
signal. The excess was found to be compatible with a DMA signal from Weakly
Interacting Massive Particles (WIMPs) in a mass range from $\sim 50$ to 100 GeV.
All properties are perfectly consistent with the WIMP being the expected neutralino
of the Minimal Supersymmetric extension of the Standard Model (MSSM) of particle
interactions \citep{deboer1}. In the analysis presented in Paper I only the
spectral shapes for the cosmic ray induced contribution and the DMA contribution to
gamma rays are used. Thus no absolute normalization of the cosmic ray induced
contribution from Galactic propagation models is needed and also the overall
experimental errors largely cancel in the relative contribution of various sky
directions which determine the distribution of dark matter. The EGRET excess indeed
traces the DM halo, as proven by reconstructing the Galactic rotation curve from it
and explaining for the first time its peculiar shape. Arguments against the DMA
interpretation of the EGRET excess concerning a too high antiproton flux are based
on simplistic propagation models with isotropic propagation \citep{bergstrom}.
These arguments can be invalidated by more realistic propagation models based on
anisotropic propagation \citep{deboer2}. Independent evidence in favor of the DMA
interpretation of the EGRET excess comes from: a) the N-body simulation of the
Canis Major dwarf galaxy as a progenitor of the Monoceros stream, which yields a
ring of Dark Matter at 13 kpc from the Galactic center \citep{penarrubia} in
perfect agreement with the EGRET excess in Paper I b)  the anomalous gas flaring in
the outer Galaxy, as was recently shown by \cite{kalberla} and c) the anomalous
change of slope in the rotation curve (see Paper I).

The flux of diffuse gamma rays has a small component from the extragalactic
background (EGB). The origin of these gamma rays are other galaxies which might
have the same production mechanisms for gamma rays as our Galaxy, or quite
different sources like Active Galactic Nuclei (AGN), quasars or blazars. Since the nature
of the extragalactic objects is unknown it is difficult to
make any predictions about the shape or the absolute value of this background
component. However, under the assumption of an isotropic extragalactic background it is
possible to determine the EGB experimentally, as done first in a pioneering paper
from \citet{sreekumar}. They determined the EGB by plotting the observed flux in
many different sky direction versus an expected Galactic flux from a Galactic Model
(GM). Extrapolating the observed flux to zero Galactic flux yields the EGB. Such an
analysis was later repeated with more advanced Galactic models \citep{exgalnew},
which yielded a lower EGB. None of the GMs used so far includes contributions from
DMA, but it is exactly DMA which may provide a large fraction of the diffuse
Galactic gamma rays, as demonstrated in Paper I and the DMA is then expected to
contribute to  the EGB as well.

Subtracting the GM contribution from the observed flux by including or excluding
DMA in the GM changes the results. E.g.
 a recent analysis of the EGB, based on  a GM without DMA
\citet{exgalnew}, could explain the high energy tail of the DMA with  a WIMP mass
of $515^{+110}_{-75}$ GeV \citep{Elsaesser}. This mass is incompatible with the
Galactic DMA signal, which requires a mass below 100 GeV as discussed above and
pointed out in a comment by \citet{comment}. It is the purpose of this paper to
show that if DMA is taken into account in the GM, the best fitted WIMP mass in the
EGB is reduced and is compatible with the WIMP mass of the Galactic excess. In
addition, if the characteristic high energy bump of the EGB is attributed to DMA,
the shape of the remaining soft part can be determined. The soft part of the EGB
was found to resemble the spectra of the observed point sources in our Galaxy.

\section{Method to determine the Extragalactic Background}\label{method}
The main concept of the EGB determination is the following (\citet{sreekumar}):
First the sky map of the diffuse gamma rays is divided into several regions to get
high and low fluxes, but the EGB is assumed to be isotropic, so each region will
have the {\it same} flux from the EGB. For these regions the measured gamma flux is
plotted against the flux of a certain GM. If the model is good, one expects a
linear dependence with a gradient of one. With an extrapolation to zero Galactic
flux from the GM one obtains the EGB from the y-axis intercept. This procedure can
be repeated for different energy intervals, thus obtaining the EGB as a function of
energy.

The uncertainties induced by the GMs can be reduced by not relying on the absolute
fluxes, but only fitting the shape of the Galactic flux predicted by the model,
thus leaving the absolute normalization of the Galactic flux a free parameter for
each sky direction. This eliminates the dependence on fluctuations in gas
densities, the interstellar radiation field and the cosmic ray densities. The total
gamma ray flux in a given direction $\Psi$ can be written as a sum of the cosmic
ray induced gamma rays interacting with the gas, as described by GMs, as well as
contributions from DMA and EGB:
\begin{equation}
  \Phi_{\mbox{\scriptsize{obs}}}(\Psi)=
  f\cdot\Phi_{\mbox{\scriptsize{GM}}}(\Psi)+g\cdot\Phi_{\mbox{\scriptsize{DM}}}(\Psi)
  +\Phi_{\mbox{\scriptsize{EGB}}} \label{fitfunction}
\end{equation}
Here $f$ and $g$ are the free normalization factors for the GM- and DM-
contribution, respectively, if only the shape is fitted. The procedure we used for
the EGB determination can be summarized as follows: first the spectral shape of the
cosmic ray induced gamma rays and the shape from DMA have to be chosen. The shape
of the GM contribution is taken from the publicly available GALPROP model
(\citet{galprop}, \citet{galprop1} and \citet{galprop2}), which is the most
detailed GM available. We used the so called conventional GALPROP model (CM), which
reproduces the locally measured fluxes of electrons and protons as well as the
secondary to primary ratio B/C. But it fails to describe the spectrum of diffuse
gamma rays, if one assumes the local CR spectra to be representative for the Galaxy.
This should be a good hypothesis, since the diffusion time is short compared to the energy
loss time. The hypothesis is supported by the fact that one can fit
the gamma ray spectrum in {\it all} sky directions with the locally measure cosmic ray spectrum.
 Since we
use only one normalization constant per sky region for the Galactic background, the
ratios of the different contributions (decay of $\pi^0$, inverse compton and
bremsstrahlung) are estimated from data, as implemented  in the GALPROP package.

The shape of the DMA is taken to correspond to a 60 GeV WIMP, as presented in Paper
I. In the EGB determination method used in this analysis the normalization factors
$f$ and $g$ of the Galactic background and the signal are obtained from a spectral
fit of the two contributions to the EGRET data. The normalization factors are
allowed to be different for each direction within an assumed uncertainty, as
expected from uncertainties in the gas densities and from a possible substructure
of the dark matter.
\begin{figure}[tbp]
  \begin{center}
    \includegraphics[width=0.4\textwidth]{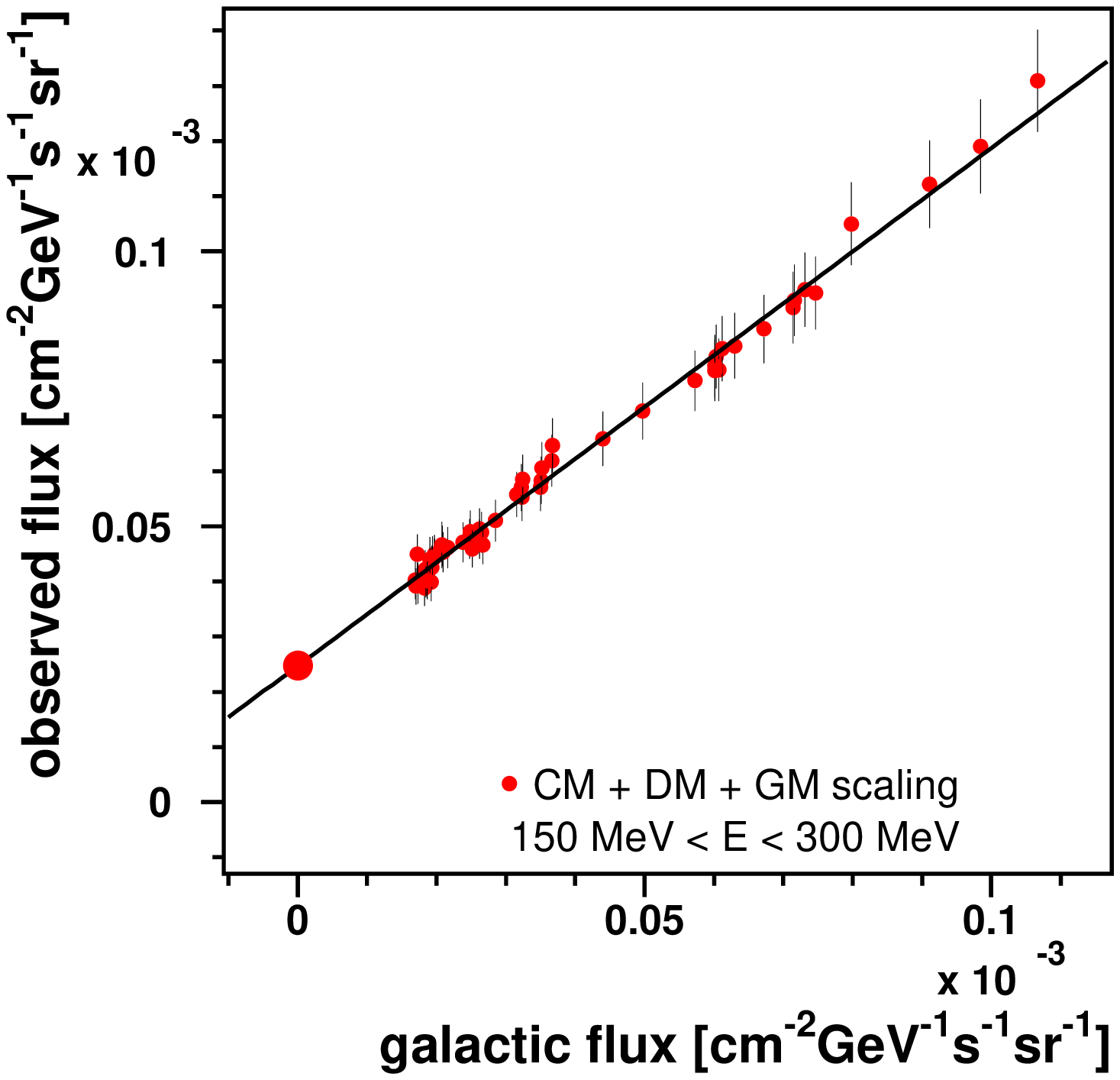}
    \includegraphics[width=0.4\textwidth]{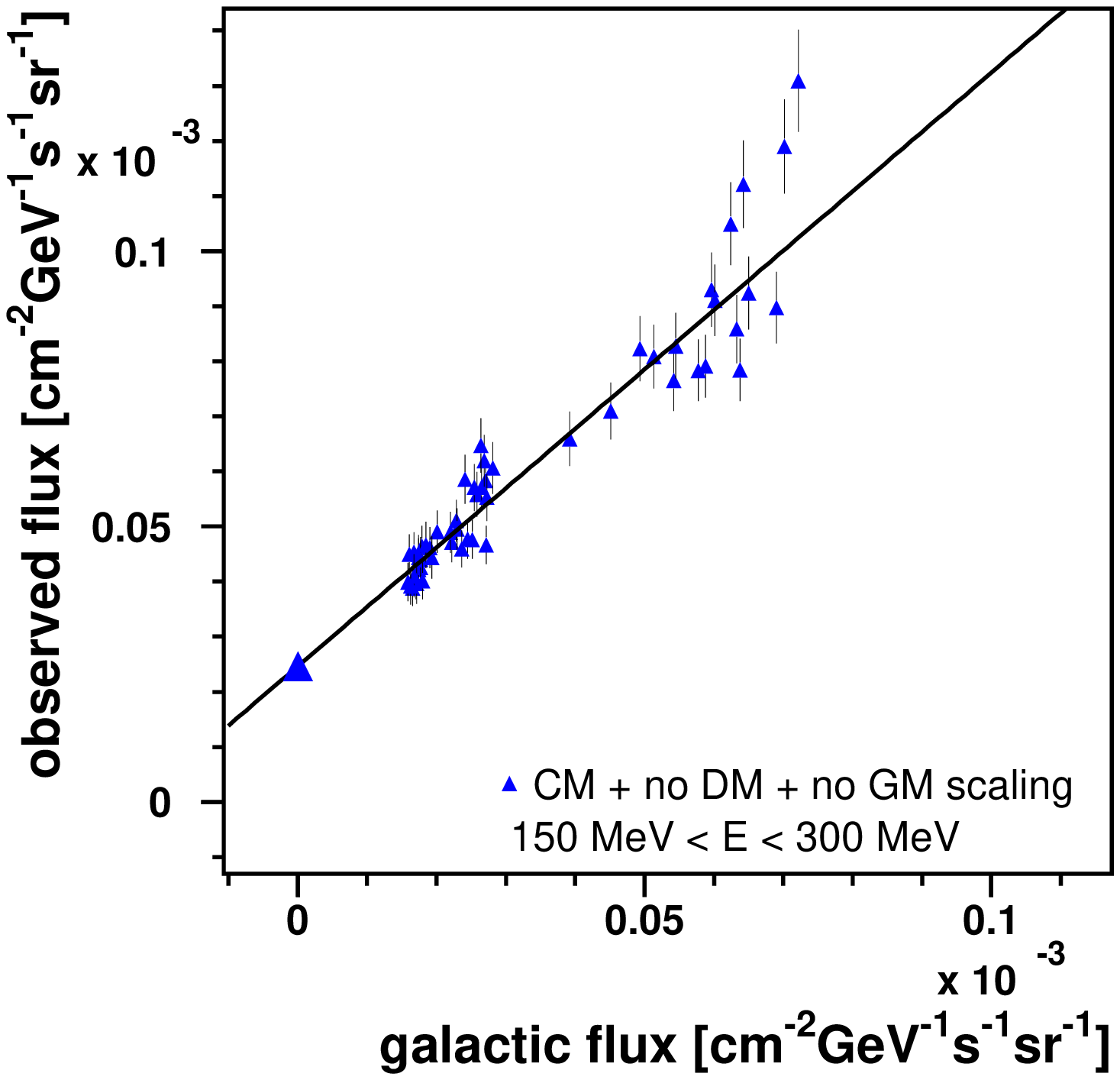}
    \includegraphics[width=0.4\textwidth]{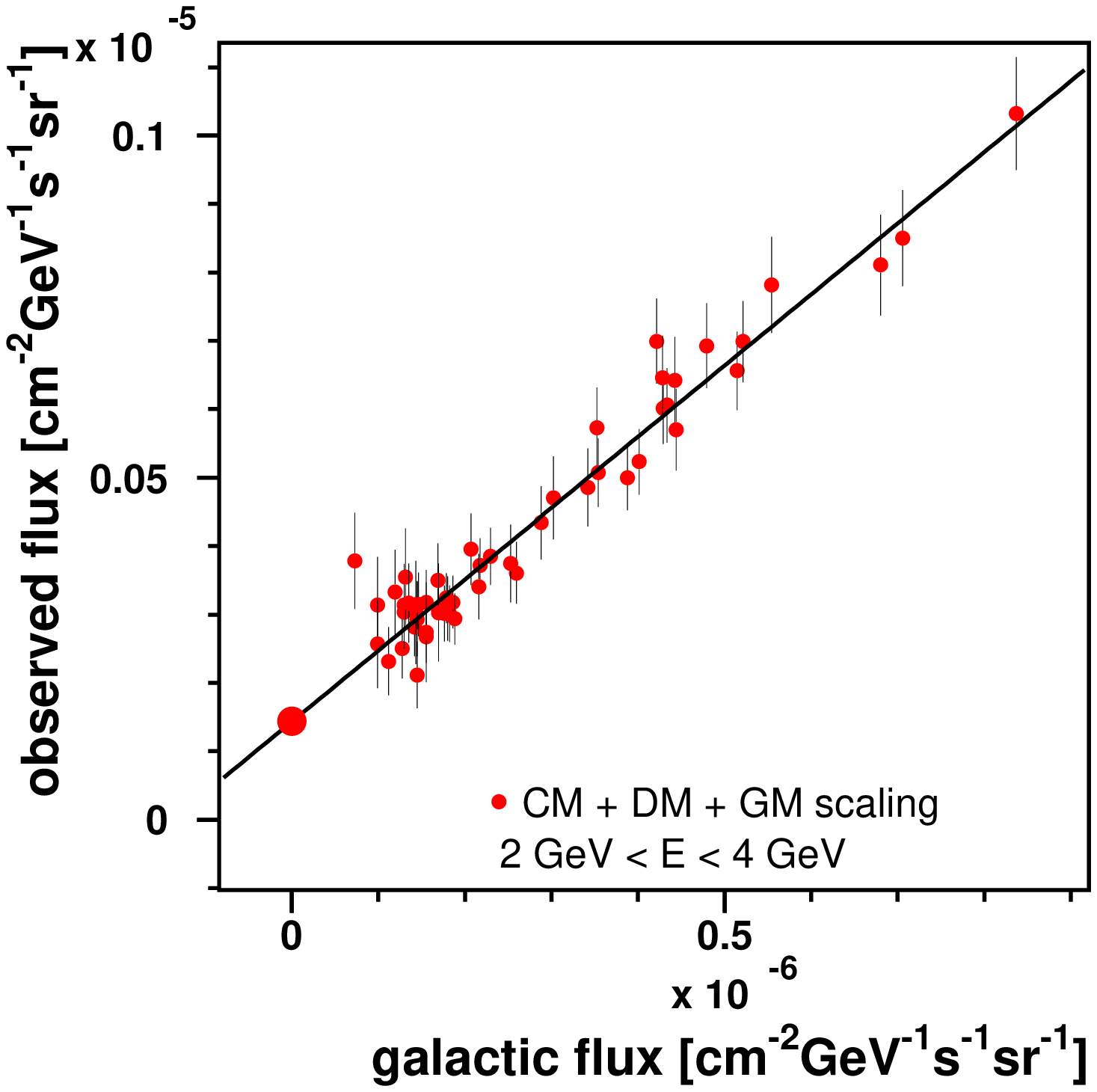}
    \includegraphics[width=0.4\textwidth]{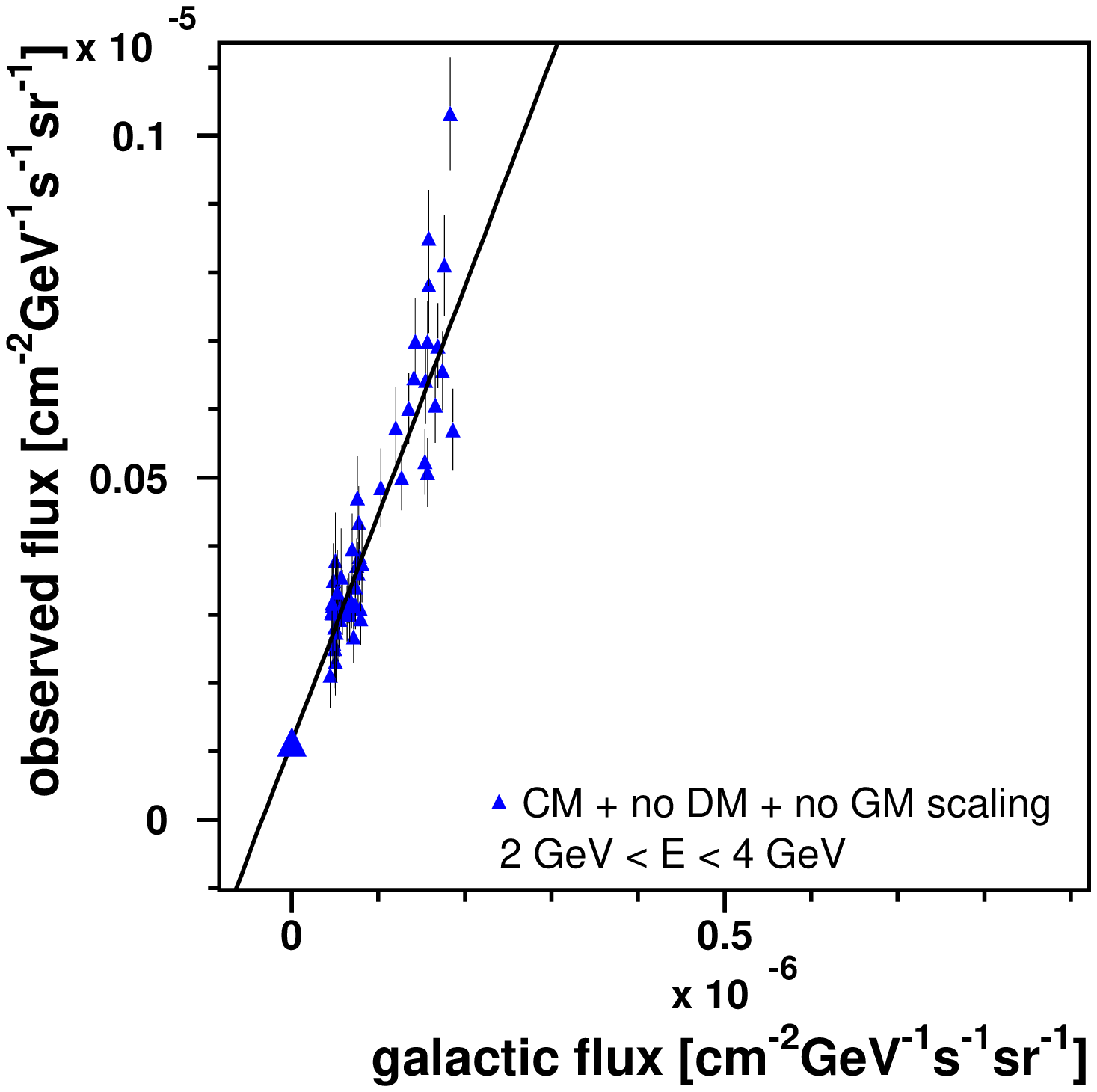}
    \includegraphics[width=0.4\textwidth]{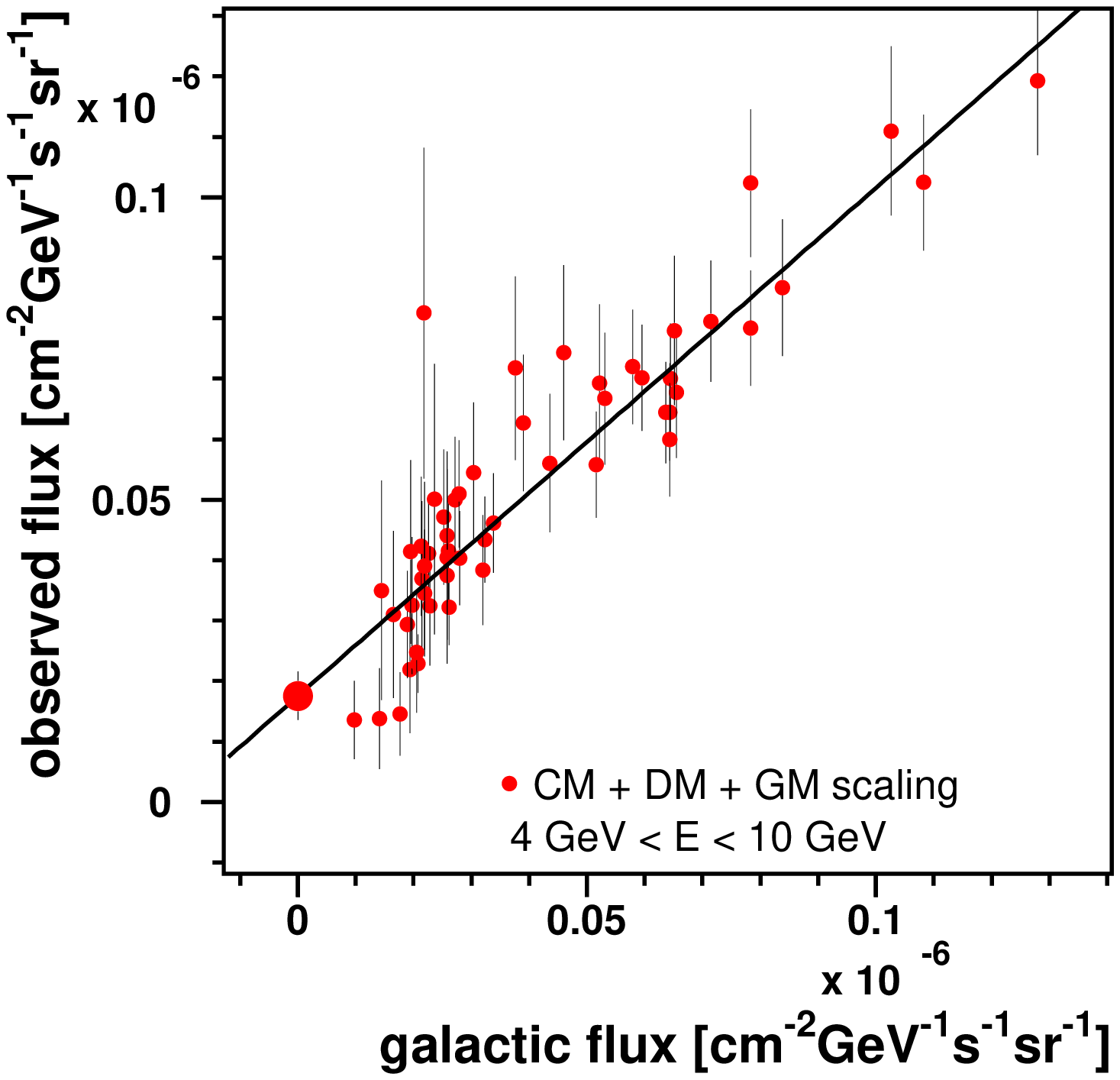}
    \includegraphics[width=0.4\textwidth]{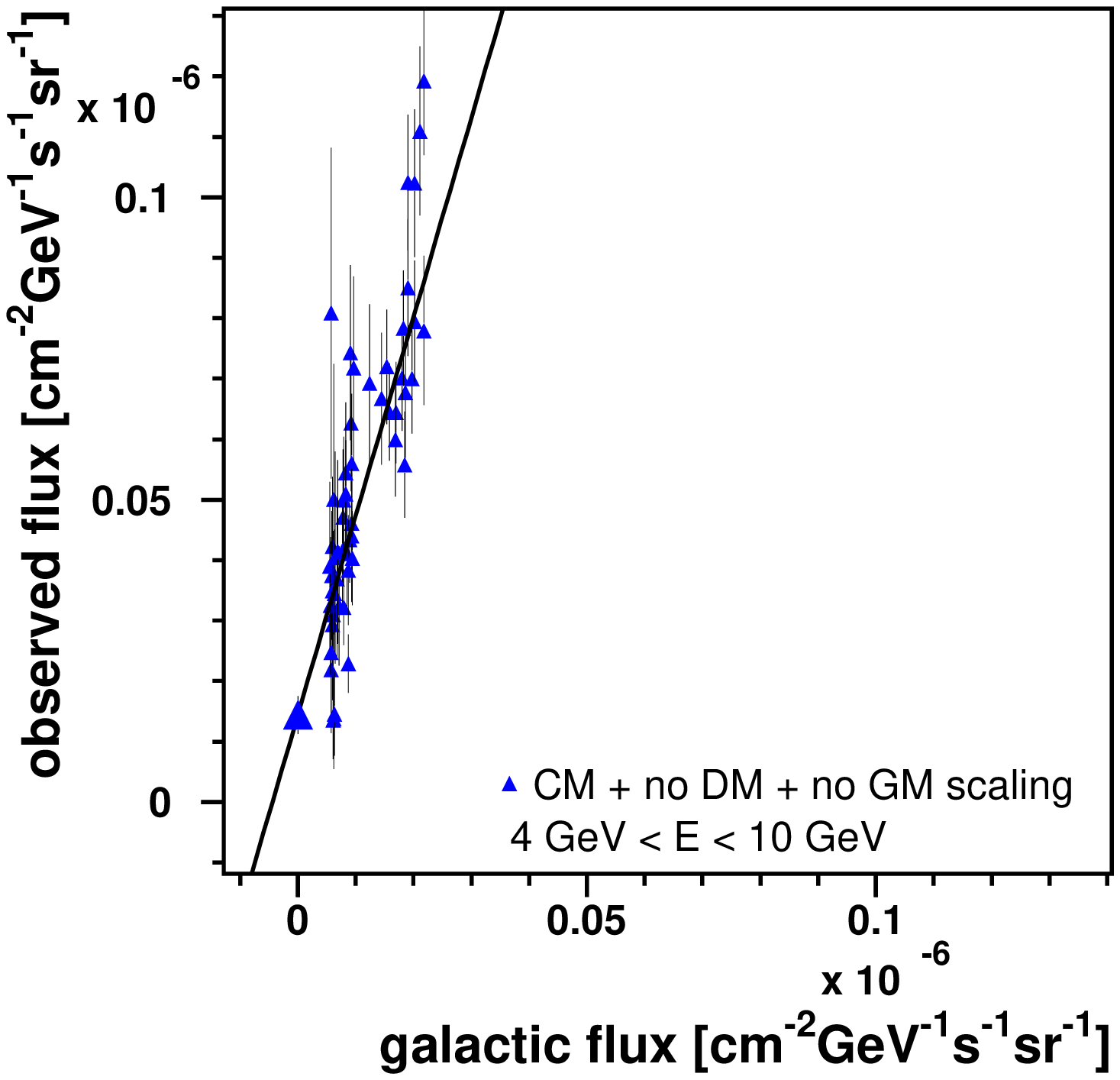}
    \caption[Determination of Extragalactic Background]{A plot of the observed flux of diffuse gamma rays (with statistical errors only) versus the Galactic flux ($=\Phi_{\mbox{\scriptsize{GM}}}+\Phi_{\mbox{\scriptsize{DM}}}$) for three of the ten EGRET energies (150 to 300 MeV (top), 2000-4000 MeV (middle) and 4 to 10 GeV (bottom). On the left hand side a DMA contribution corresponding to a WIMP mass of $\sim 60$ GeV was taken into account and the GM normalization was left as a free parameter, while on the right hand side the EGB was determined in the simplest way, i.e. the Galactic flux is directly taken from GALPROP without scaling and without DMA contribution. Note the strong change of slope at high energies caused by the DMA contribution. The extrapolated point (large symbol) at the Galactic flux $\Phi_{\mbox{\scriptsize{GM}}}+\Phi_{\mbox{\scriptsize{DM}}}=0$ is the EGB.}
    \label{exgal}
  \end{center}
\end{figure}

The scaling parameters $f$ and $g$ of the GM and the DMA contribution depend on the
EGB, since a larger extragalactic component leads to a smaller galactic one and
vice versa. Therefore one has to determine the EGB iteratively. The following
algorithm was used:
\begin{enumerate}
\item Set the extragalactic background to zero.
\item
Determine the direction dependent scaling factors $f^{i,j}$ and $g^{i,j}$ for
$i_{max}\times j_{max}=18\times 3 =54$ independent sky regions from a fit to the
spectral shape of the observed gamma ray flux in the region with indices $i,j$ , as
done in Paper I. The indices $i$ and $j$ denote the longitude and latitude
interval, respectively. In this paper we use 18 equidistant intervals in galactic
longitude and 3 equidistant intervals from 10 to 90 degrees. We also tried 6
intervals for the latitude to take positive and negative values separately. The
Galactic plane is excluded to eliminate effects from spiral arms and other DM
structures in the disc, as observed in Paper I. The scaling factors $f^{i,j}$ are
constrained to a value of $f=1$ within the normalization uncertainty $\sigma_f$ by
adding a term $(f-1)^2/\sigma_f^2$  to the $\chi^2$.
  Similarly
the scaling of the DMA is constrained by adding a term $(g-1)^2/\sigma_g^2$.
Numerical values of these scaling errors will be discussed later. The scaling
factors $f^{i,j}$ and $g^{i,j}$  have been determined from the fit to the spectral
shape of the gamma rays in the region $(i,j)$, as in Paper I.
\item Choose an energy interval. The observed flux is plotted against the
Galactic flux
$f\cdot\Phi_{\mbox{\scriptsize{GM}}}+g\cdot\Phi_{\mbox{\scriptsize{DM}}}$ for the
sky regions introduced earlier. Clearly
$\Phi_{\mbox{\scriptsize{EGB}}}=\Phi_{\mbox{\scriptsize{obs}}}$ for
$f\cdot\Phi_{\mbox{\scriptsize{GM}}}+g\cdot\Phi_{\mbox{\scriptsize{DM}}}=0$, i.e.
the EGB can be obtained from the y-axis intercept of a linear fit (Eq.
\ref{fitfunction}) through these points. Since the fit function contains the
unknown $\Phi_{\mbox{\scriptsize{EGB}}}$, the final solution can only be found by
an iterative procedure, described by the next steps.
\item Go back to step 3 and
repeat the procedure for the next energy interval of gamma rays in order to obtain
the complete EGB spectrum.
\item Go back to step 2, unless the EGB has not changed
within a given accuracy.
\end{enumerate}

\section{Results}\label{results}

\subsection{EGB determination}

In Fig. \ref{exgal} the linear fits of the correlation between the observed flux
and the GM flux are shown for 3 different energy ranges following the description
of the previous section. For comparison also the EGB using GALPROP without GM
scaling and without DMA contribution is shown on the right hand side. Clearly the
slopes and offsets representing the EGB are significantly different. For a model
including DMA the slope is close to one, as expected for a correct model.
Nevertheless the model without DM has almost the same intercepts and therefore the
same EGB. At high energies the difference of the extrapolation originates from the
DMA contribution and the GM scaling, while at lower energies only the GM scaling
effects the result.

\begin{figure}[tbp]
  \begin{center}
    \includegraphics[width=0.32\textwidth]{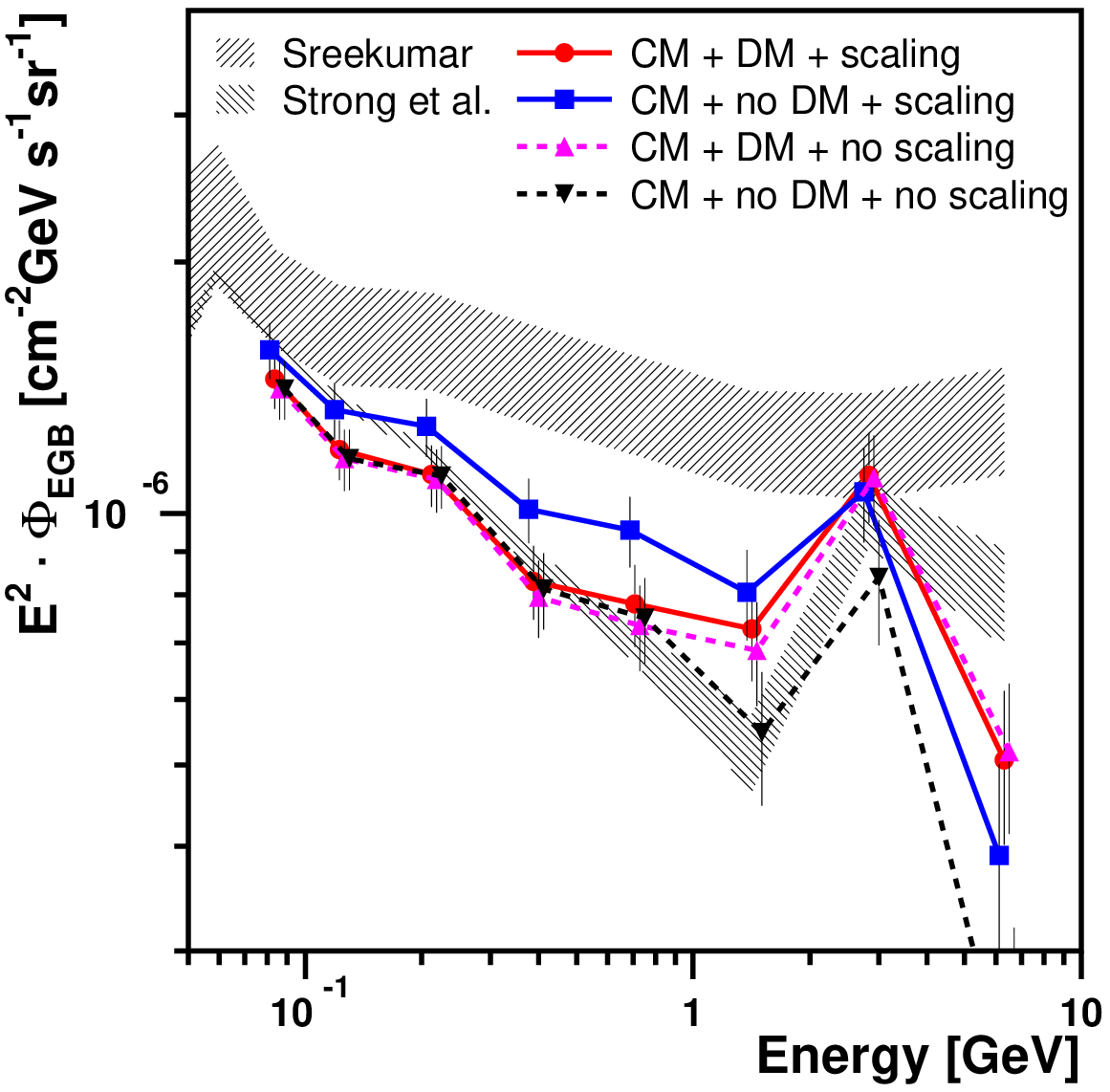}
    \includegraphics[width=0.32\textwidth]{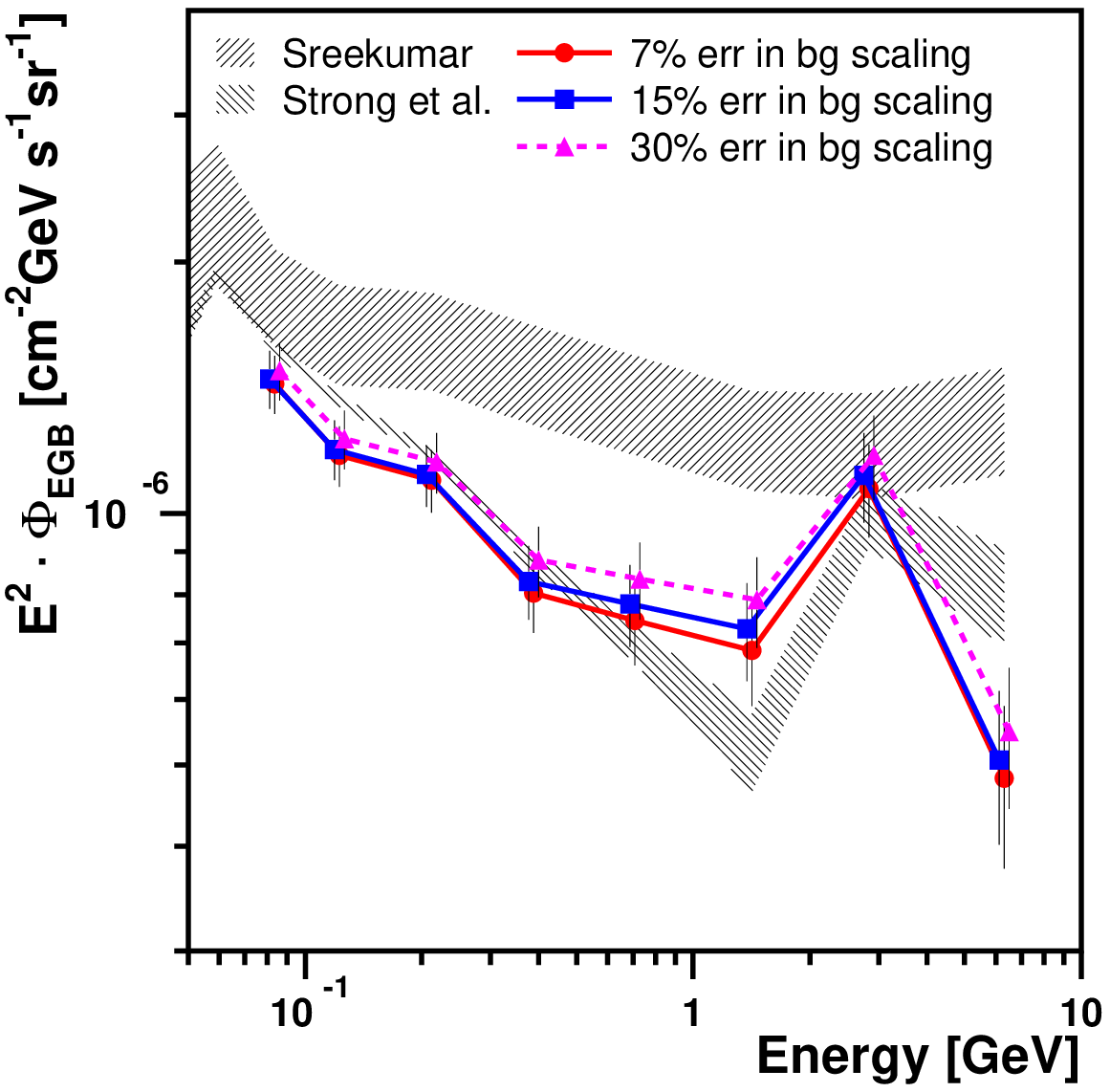}
    \includegraphics[width=0.32\textwidth]{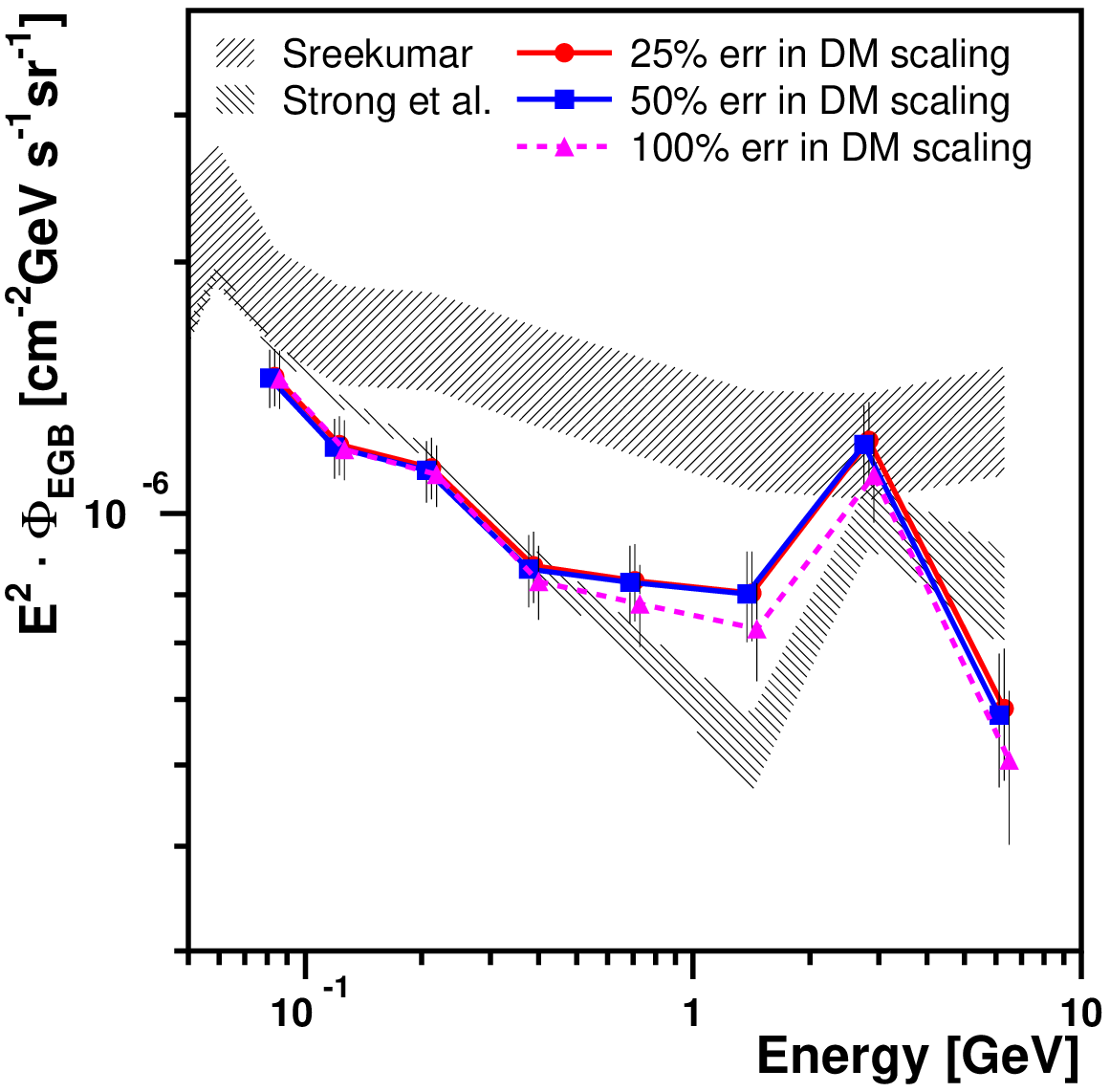}
    \caption[Spectrum of Extragalactic Background]{Left: the EGB with  four different
    curves corresponding to different methods (with/without DMA contribution and
    with/without GM scaling. Only statistical errors are plotted.
    In the middle and right hand panel the dependence
    of the EGBs on the scaling errors $\sigma_f$ and $\sigma_g$ are shown.
    The two shaded bands are the EGB determined by \citet{sreekumar}
    (highest band,  including systematic errors) and \citet{exgalnew} (lower band, only statistical errors).} \label{exgalspec}
  \end{center}
\end{figure}

After repeating these linear fits for the 8 highest EGRET energy intervals (above
70 MeV) one obtains the spectral shape of the extragalactic background, as shown in
Fig. \ref{exgalspec}. The EGBs for the CM are calculated  with the four different
combinations: with/without GM scaling for each sky map region and with/without DMA
signal.

The following points are noteworthy:
\begin{itemize}
\item A comparison with previous EGB determinations, indicated by
the hatched areas (upper one from \citet{sreekumar}, lower one from
\citet{exgalnew}) shows that our results  are in reasonable agreement with the ones
from \citet{exgalnew}. This is not surprising since the lower hatched area has also
been determined with the GALPROP package, while the upper one is based on a simpler
model with an underestimated inverse Compton contribution.  The effect of the
scaling of the GM contribution increases the EGB, while the DMA decreases the EGB
below 3 GeV and enhances it above 3 GeV  (see left panel of Fig. \ref{exgalspec}).
Therefore the region around 3 GeV is quite stable. The net result is a decrease of
the high energy tail and enhancing the structure around 3 GeV. Note that both
previous determinations have not been done with a region dependent GM scaling, but
a scaling for each energy interval to take care of the imperfect description of the
spectral gamma ray shape. This scaling is taken to be the same for all regions and
leads to the difference between the methods.
\item In the
middle panel of Fig. \ref{exgalspec} the EGB is shown for three values of the
normalization error, namely $\sigma_f-0.07, 0.15, 0.3$, respectively.  The EGB
increases continuously for an increasing uncertainty of the scaling. This effect is
due to the correlation between the DMA signal, the GM and the EGB, which can be
understood as follows: let us assume that the spectral fit of the diffuse gamma
rays prefers a smaller GM contribution, i.e. a smaller $f^{i,j}$. This shifts the
fitted line in a correlation plot like Fig. \ref{exgal} to the left which  leads to
a larger y-axis intercept, i.e. a larger EGB. Consequently in the next iteration
step a even smaller GM contribution is needed which causes the algorithm to
converge at a larger EGB. A similar effect happens with  the uncertainty in the DMA
scaling, but with an opposite sign, as shown in the right panel and  in Table
\ref{EGBvalues}. The overall spectral shape of the EGB is only weakly dependent on
these systematic errors, so the characteristic bump above 1 GeV is hardly affected.

\end{itemize}

\begin{table}[tbp]
  \begin{center}
    \begin{tabular}{|c|c|c|c|c|c|c|}
      \hline
      Energy [MeV] & $E^2\cdot \Phi_{\rm EGB}$ [GeV cm$^{-2}$s$^{-1}$sr$^{-1}$] & $\Phi_{\rm EGB}
      $ [GeV$^{-1}$ cm$^{-2}$s$^{-1}$sr$^{-1}$] & $\sigma_{EGRET}$ & $\sigma_{GM}$ & $\sigma_{DM}$  & $\sigma_{tot}$\\
      \hline
      70 \ldots 100     & 1.45(5) $\times 10^{-6}$  & 2.07(8) $\times 10^{-4}$  & 0.7\%  & 1.1\% & 1.0\% & 7.2\%\\
      100 \ldots 150    & 1.19(5) $\times 10^{-6}$  & 7.96(34) $\times 10^{-5}$ & 0.8\%  & 0.6\% & 0.7\% & 7.1\%\\
      150 \ldots 300    & 1.11(5) $\times 10^{-6}$  & 2.47(12) $\times 10^{-5}$ & 1.7\%  & 0.4\% & 0.3\% & 7.2\%\\
      300 \ldots 500    & 0.83(6) $\times 10^{-6}$  & 5.53(41) $\times 10^{-6}$ & 3.6\%  & 1.7\% & 0.8\% & 8.1\%\\
      500 \ldots 1000   & 0.78(7) $\times 10^{-6}$  & 1.56(14) $\times 10^{-6}$ & 1.0\%  & 2.8\% & 1.8\% & 7.8\%\\
      1000 \ldots 2000  & 0.73(8) $\times 10^{-6}$  & 3.64(42) $\times 10^{-7}$ & 4.2\%  & 4.1\% & 3.5\% & 9.8\%\\
      2000 \ldots 4000  & 1.11(11) $\times 10^{-6}$ & 1.39(14) $\times 10^{-7}$ & 8.6\%  & 1.7\% & 3.4\% & 11.7\%\\
      4000 \ldots 10000 & 0.51(10) $\times 10^{-6}$ & 1.26(25) $\times 10^{-8}$ & 12.7\% & 3.6\% & 5.7\% & 16.1\%\\
      \hline
    \end{tabular}
  \end{center}
  \caption{The EGB flux ($E^2\Phi_{\rm EGB}$ and $\Phi_{\rm EGB}$) for different energies
  with the statistical errors in brackets. For the energy $E$ of
  an interval the geometric average $\sqrt{E_{\rm max}\cdot E_{\rm min}}$ is taken. In
  columns 4 to 6 the systematic uncertainties on the spectral shape  from the data, GM uncertainty and
  DM uncertainty
  are shown.  They
  were determined by half of the variation of the EGB caused by changing the EGRET
  data by $\pm 15\%$, which is the estimated EGRET absolute normalization error and
  by varying the GM and the DM scaling errors as indicated in Fig. \ref{exgalspec}. Note that
  increasing the EGRET data will also increase the prediction of the GM, since only the
  shape is fitted, so the effect of the normalization uncertainty is reduced. The total
  systematic uncertainty in the last column includes a point to point error of
  the EGRET experiment which is taken from Paper I to be 7\%.  All errors are added in quadrature.} \label{EGBvalues}
\end{table}

\subsection{Anisotropy of the EGB}
\begin{figure}[tbp]
  \begin{center}
    \includegraphics[width=0.32\textwidth]{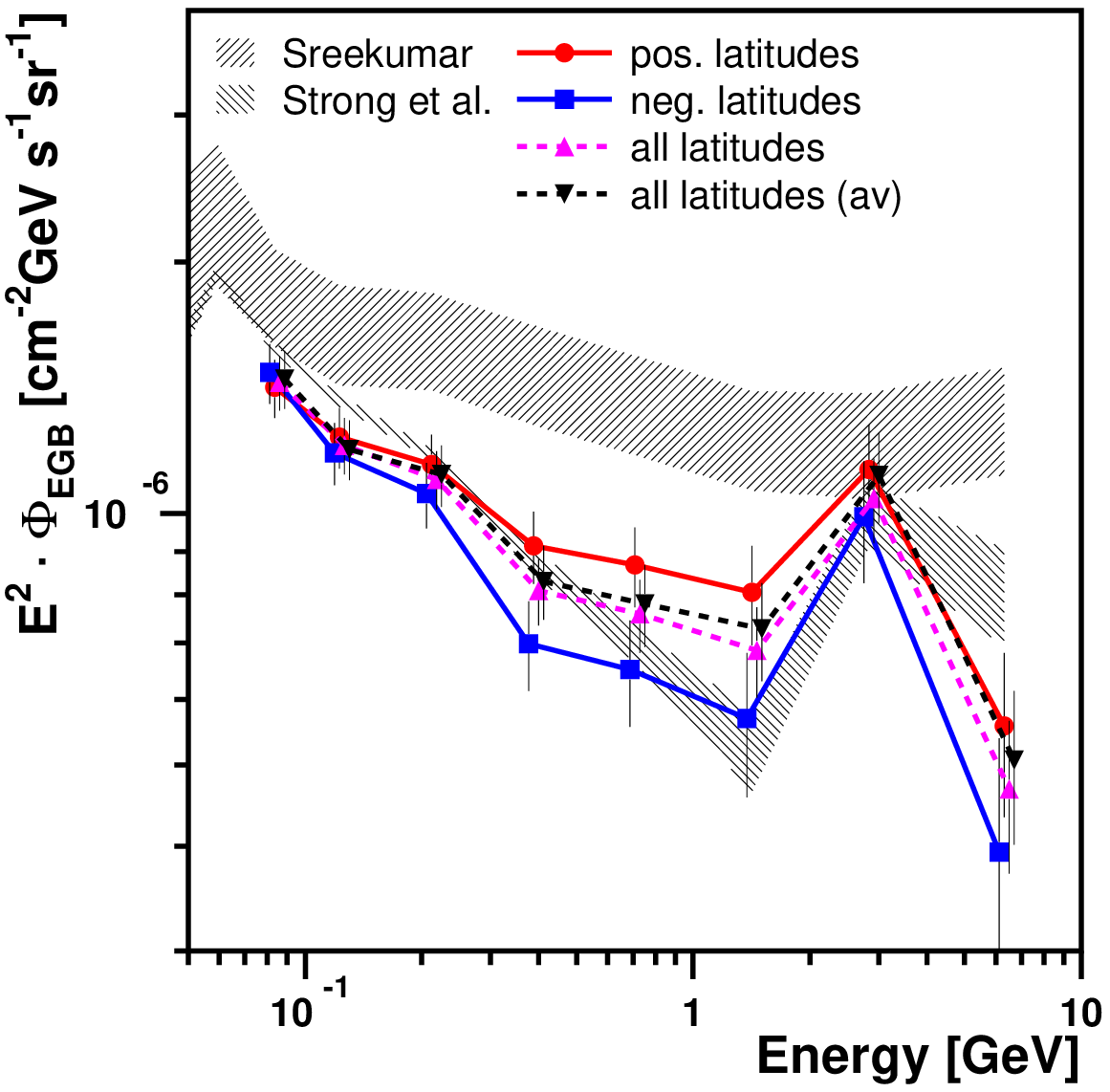}
    \includegraphics[width=0.32\textwidth]{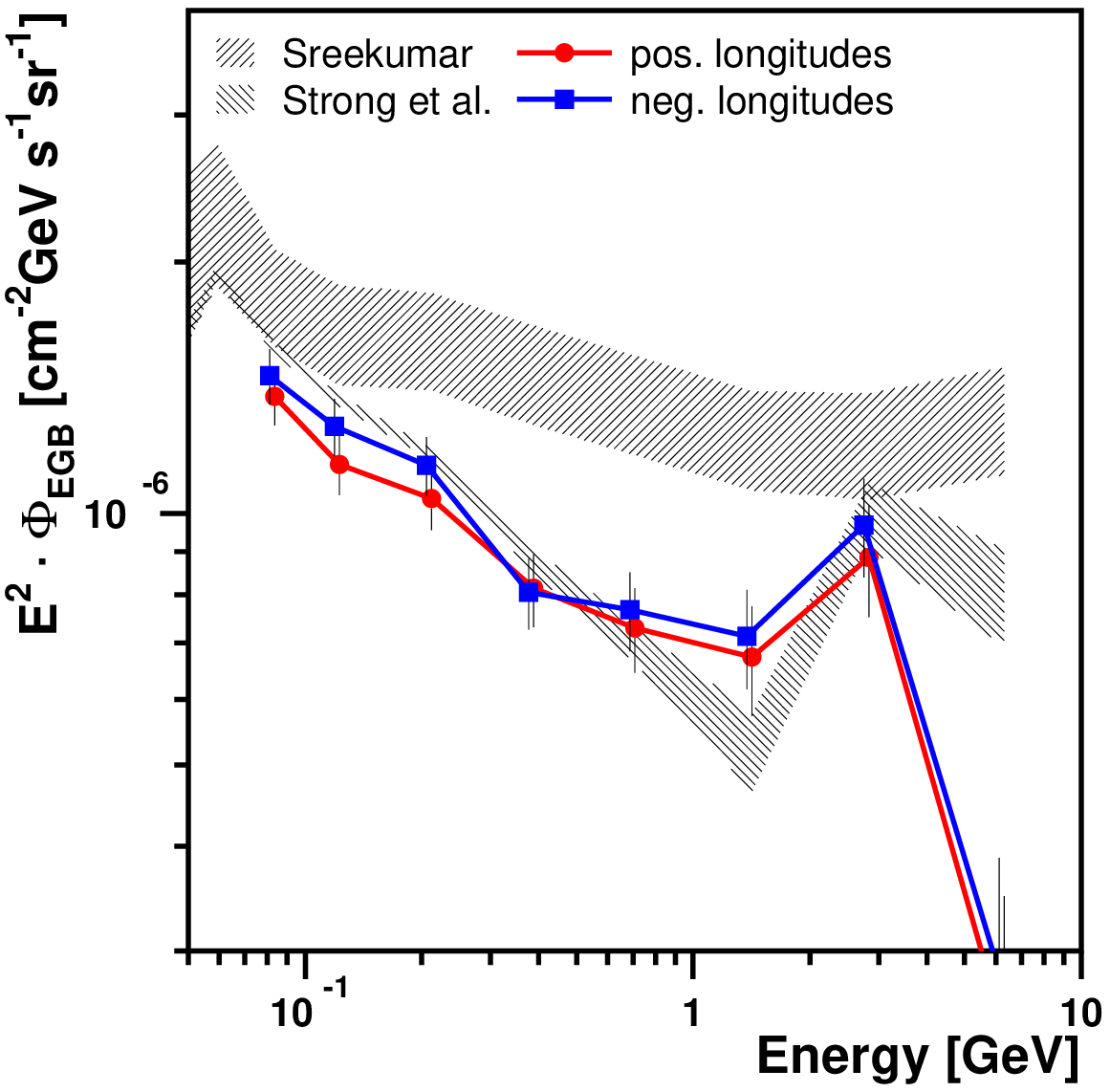}
    \caption{Spectrum of the extragalactic background determined for the
   conventional model with GM scaling and DMA for different hemispheres. The
   systematic point to point error is not included since it is common to all sky
   regions, so only the statistical errors are plotted.}
   \label{exgalspec_hemisphere}
  \end{center}
\end{figure}

The EGB in the previous section was assumed to be isotropic. To test this
hypothesis the EGB has been determined for various subsections of the sky. The
Galactic plane (latitudes smaller than 10$^\circ$) has been excluded in all cases,
since in this region the EGB is much lower than the Galactic gamma rays and the
diffuse gamma rays strongly depend on inhomogeneous structures, both in the
baryonic and dark matter. In Fig. \ref{exgalspec_hemisphere} the EGB is shown for
the hemispheres with positive or negative latitude as well as for positive or
negative longitudes. The different hemispheres for positive and negative longitudes
show good agreement with each other, while the hemispheres with positive and
negative latitudes have a difference of $\sim$ 15\%. A similar anisotropy was also
observed by \citet{sreekumar2} and it shows that the EGB is not perfectly
isotropic, if one assumes it is not an experimental effect. This is not excluded, since the
detector efficiency changed in time and the exposure times were not equal for all sky directions.
 An alternative reason for the observed anisotropy could be the fact, that the earth
is not located exactly in the galactic plane but a few tens of pc above the plane.
Or equivalently the Milky Way baryonic and/or DM contribution is not perfect
symmetric with respect to the plane, but in the GM it is assumed to be symmetric.
Therefore it is important to average the regions for positive and negative
latitudes. It should be pointed out that for both hemispheres the EGB spectra show
the characteristic feature of the high energy bump.

\subsection{Contribution of DMA in the EGB}

The bump in the EGB at 3 GeV  is indicative of an extragalactic DMA contribution.
This characteristic feature  does not depend on any of the uncertainties discussed
in the previous sections.  If this feature is fitted with a DMA contribution for a
60 GeV Wimp, as needed to describe the Galactic EGRET excess, then the remaining
part of the spectrum is the contribution of  all extragalactic sources except the
DMA contribution. This remaining part can be parametrized by a double power law,
typically used for point sources:
\begin{equation}
  E^2\cdot\Phi_{\mbox{\scriptsize{EGB}}}={2\Phi_0\over
  \left( E\over E_0 \right)^{-\gamma_1}+
  \left( E\over E_0 \right)^{-\gamma_2}} \label{parametrization}
\end{equation}

\begin{figure}[tbp]
  \begin{center}
    \includegraphics[width=0.32\textwidth]{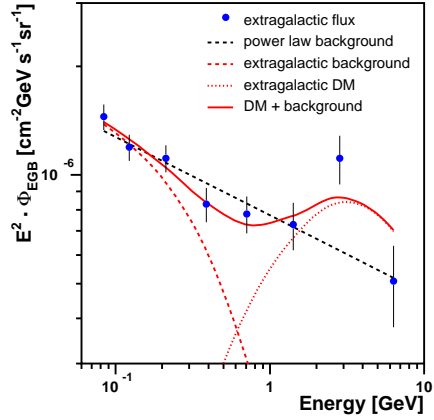}
    \caption{Fits to the extragalactic background with a single power law (dashed black)
    and a double power law (dashed, red) plus a DMA contribution (dotted, red). In
    this plot the complete error is included.}
    \label{exgalfit}
  \end{center}
\end{figure}

The best fit parameters of a combined fit of DMA plus EGB$_{non-DM}$ are given in
Table \ref{FITvalues} and the fits are compared with the data in Fig.
\ref{exgalfit}. There is a strong correlation between the two slopes in the double
power law and the transition energy $E_0$, i.e. the EGB at low energies can be
fitted either by the contribution below or above the break. So we fixed the
transition energy $E_0$ to 0.5 GeV which hardly changes the total $\chi^2$. The
model including DMA has a larger probability than the model without DM (using a
single power law, see Table \ref{FITvalues}),
 because of the two highest energy points. A WIMP mass of 60 GeV provides certainly
a possibility to explain the bump in the EGB and is compatible with the excess in
the Galactic data
 in contrast to the WIMP mass determination by \citep{Elsaesser},
as discussed in the introduction.
 Note that the statistical significance for the single high point
is not overwhelming, as shown by the rather small difference in probabilities of the single and double
power law in Table  \ref{FITvalues}. Therefore other explanations, like additional contributions
from AGN's or blazars are certainly not excluded.  However, the fact that the bump just occurs at the
energy interval, where also the Galactic excess has a maximum, strengthens the DMA interpretation.
In this case the dominant contribution to the EGB flux of diffuse gamma rays at
high energies originated from DMA ( see Fig. \ref{exgalfit}), just like it is the case for the Galactic flux,
which is also dominated by DMA for energies above 2 GeV.
The remaining soft contribution of the EGB resembles the  spectra of the many point sources
in our Galaxy, as shown in the Appendix of Paper I.

\begin{table}[tbp]
  \begin{center}
    \begin{tabular}{|c|c|c|}
      \hline
      Parameter & single power & double power \\
      \hline
      $\Phi_0$ [GeV cm$^{-2}$s$^{-1}$sr$^{-1}$] & $(4.5\pm 0.2)\cdot 10^{-7}$ & $(5.3\pm 0.6)\cdot 10^{-7}$\\
      $E_0 [\rm GeV]$                               & 0.5                         & 0.5\\
      $\gamma_1$                                & $-0.21 \pm 0.04$            & $-0.19 \pm 0.09$\\
      $\gamma_2$                                & -                           & $-1.74 \pm 0.53$\\
      \hline
      $\chi^2/$d.o.f.                           & 11.7/6                      & 5.7/4 \\
      probability                               & $6.9\cdot 10^{-2}$          & $22.4\cdot 10^{-2}$ \\
      \hline
    \end{tabular}
  \end{center}
  \caption{Parameters of the fit to the shape of the EGB  with a single power law and a double power law
  (see Eq. \ref{parametrization}); in the latter case the DMA was
  included in the fit  using the shape of the DMA spectrum of Paper I.
  } \label{FITvalues}
\end{table}

\section{Conclusion}

A new iterative method for the determination of the isotropic EGB has been
presented, relying mainly on the spectral {\it shape} of the gamma rays from a
Galactic Model. The influence of the Galactic DMA on the EGB was studied as well.
The newly determined EGB shows a characteristic bump around 3 GeV, i.e. at the same
position as the Galactic EGRET excess. Therefore this bump can be fitted with the
same WIMP mass as determined in Paper I from the Galactic excess, which is much
lower than the WIMP mass around 500 GeV found by \citet{Elsaesser}. This bump is
insensitive to all possible uncertainties considered and is therefore real,
although its statistical significance is less than $2\sigma$, since a single power
law gives already a probability of 7\%. If nevertheless a DMA contribution is
fitted, the EGB flux above 1 GeV is dominated by the  {\it extragalactic} DMA, as
shown in Fig. \ref{exgalfit} and the remaining  soft contribution in the EGB can be
described by a  power law reminiscent of the spectra of the typical point
sources in our Galaxy (see Paper I).

\acknowledgements We thank  O. Reimer for a meticulous reading of the manuscript and
healthy criticism.  This work was supported by BMBF (Bundesministerium f\"ur Bildung und
Forschung) via the DLR (Deutsches Zentrum f\"ur Luft- und Raumfahrt).

\bibliographystyle{aa}

\begin{thebibliography}{199}
\bibitem[De Boer et al.(2005)]{deboer} de Boer, W., Sander, C., Zhukov, V., Gladyshev, A.V. and Kazakov, D.I., 2005,
         A\&A 444, 51; arXiv:astro-ph/0508617.
\bibitem[De Boer et al.(2006)]{deboer1} de Boer, W., Sander, C., Zhukov, V., Gladyshev, A.V. and Kazakov, D.I., 2006,
        Phys.\ Lett.\ B 636, 13; arXiv:hep-ph/0511154.
\bibitem[Bergstr$\rm\ddot o$m et al. (2006)]{bergstrom} Bergstr$\rm\ddot o$m, L., Edsj$\rm\ddot o$, J.,
 Gustafsson, M., and Salati, P.,
  JCAP { 0605}, 006,
  [arXiv:astro-ph/0602632].
\bibitem[De Boer et al. (2006)]{deboer2} de Boer, W., Gebauer, I., Sander, C., Weber, M.,
 and Zhukov, V., to be published in the Proc. of the 7th UCLA Symposium on Sources and
 Detection of Dark Matter and Dark Energy in the Universe, Marina del Rey, 2006,
  arXiv:astro-ph/0612462.
\bibitem[De Boer et al.(2005a)]{comment} de Boer, W., Sander, C., Zhukov, V., Gladyshev, A.V. and Kazakov, D.I., 2005,
            Phys.\ Rev.\ Lett.\ 95, 209001.
\bibitem[Els\"asser et al. (2005)]{Elsaesser} Els\"asser, D., Mannheim, K. 2005,
         Phys.\ Rev.\ Lett.\  94, 171302.
\bibitem[Esposito et al.(1999)]{egret_cal1} Esposito, J.A. et al. 1999,
         ApJ 123,  203.
\bibitem[Hunter et al.(1997)]{hunter} Hunter, S.D. et al. 1997,
         ApJ 481, 205.
\bibitem[Kalberla et al.(2007)]{kalberla} Kalberla, P.M.W.,  Dedes, L.,  Kerp, J., and  Haud, U. 2007,
"Dark matter in the Milky Way II. the HI gas distribution as a tracer of the gravitational potential",
  submitted to Astrophys.\ J.
\bibitem[Moskalenko, Strong \& Reimer(1998)]{galprop1} Moskalenko, I.V., Strong, A.W. and Reimer, O. 1998,
         A\&A 338, L75; arXiv:astro-ph/9808084
\bibitem[Penarrubia et al. (2005)]{penarrubia}
  Penarrubia, J. {\it et al.,}  [SDSS Collaboration],
  Astrophys.\ J.\  { 626}, 128.
  [arXiv:astro-ph/0410448].
\bibitem[Sreekumar et al.(1997)]{sreekumar2} Sreekumar, P. and Stecker, F.~W. and Kappadath, S.~C. 1997,
         AIP Conf. Proc. 410: Proceedings of the Fourth Compton Symposium, 344; arXiv:astro-ph/9709258.
\bibitem[Sreekumar et al.(1998)]{sreekumar} Sreekumar, P. et al. [EGRET Collaboration] 1998,
         ApJ 494, 523; arXiv:astro-ph/9709257.
\bibitem[Strong \& Moskalenko(1998)]{galprop} Strong, A.W. and Moskalenko, I.V. 1998,
         ApJ 509, 212; arXiv:astro-ph/9807150.
\bibitem[Strong, Moskalenko \& Reimer(2000)]{galprop2} Strong, A.W., Moskalenko I.V. and Reimer O. 2000,
         ApJ 537, 763 [Erratum-ibid. 541, 1109]; arXiv:astro-ph/9811296; Details on the latest GALPROP versions can be found at http://www.mpe.mpg.de/~aws/propagate.html.
\bibitem[Strong, Moskalenko \& Reimer(2004)]{exgalnew} Strong, A.W., Moskalenko, I.V. and Reimer, O.,
         ApJ 613, 956; arXiv:astro-ph/0405441.
\bibitem[Strong, Moskalenko \& Reimer(2004)]{optimized} Strong, A.W., Moskalenko, I.V. and Reimer, O. 2004,
         ApJ 613, 962; arXiv:astro-ph/0406254.
\bibitem[Thompson et al.(1987)]{egret_cal} Thompson D.J. et al. 1987,
         IEEE Trans. Nucl. Sci. 34, 36.
%
\end{thebibliography}

\end{document}